\documentclass[aps,pra,reprint,superscriptaddress]{revtex4-1}
\usepackage{graphicx}
\usepackage{amssymb}
\usepackage{amsmath}
\usepackage{siunitx}
\usepackage{color}

\DeclareSIUnit\Molar{\textsc{m}}

\begin{document}
	
\preprint{AIP/123-QED}
\title{Two step micro-rheological behavior in a viscoelastic fluid}


\author{Rohit Jain\normalfont\textsuperscript{$\dagger$,}}
\email{rohit.jain@theorie.physik.uni-goettingen.de}
\affiliation{Institute for Theoretical Physics, Georg-August Universit\"{a}t G\"{o}ttingen, 37073 G\"{o}ttingen, Germany}

\author{F\'{e}lix Ginot}
\altaffiliation{Both authors contributed equally to this work.}
\affiliation{Fachbereich Physik, Universit\"{a}t Konstanz, 78457 Konstanz, Germany}

\author{Johannes Berner}
\affiliation{Fachbereich Physik, Universit\"{a}t Konstanz, 78457 Konstanz, Germany}

\author{Clemens Bechinger}
\affiliation{Fachbereich Physik, Universit\"{a}t Konstanz, 78457 Konstanz, Germany}

\author{Matthias Krüger}
\affiliation{Institute for Theoretical Physics, Georg-August Universit\"{a}t G\"{o}ttingen, 37073 G\"{o}ttingen, Germany}



\date{\today}

\begin{abstract}
We perform micro-rheological experiments with a colloidal bead driven through a viscoelastic worm-like micellar fluid and observe two distinctive shear thinning regimes, each of them displaying a Newtonian-like plateau. The shear thinning behavior at larger velocities is in qualitative agreement with macroscopic rheological experiments. The second process, observed at Weissenberg numbers as small as a few percent, appears to have no analog in macro rheological findings. A simple model introduced earlier captures the observed behavior, and implies that the two shear thinning processes correspond to two different length scales in the fluid. This model also reproduces oscillations which have been observed in this system previously. While the system under macro-shear seems to be near equilibrium for shear rates in the regime of the intermediate Newtonian-like plateau, the one under micro-shear is thus still far from it. The analysis suggests the existence of a length scale of a few micrometres, the nature of which remains elusive. 
\end{abstract}

\maketitle

 

\section{Introduction}\label{sec:INTRO}
Stochastic processes are of general importance, from a fundamental point of view but also regarding technical and biological applications. As a result, they have been the subject of intense research over the past years. A simple example is Brownian motion which has been intensively studied, both in equilibrium and in presence of various types of external driving, in experiments and theoretically~\cite{Seifert2012-wr, Sekimoto1998-cb, Dhont1996-vb}. In particular in case of entirely viscous, i.e. Newtonian solvents the treatment of such processes is rather straightforward because it can be described within the framework of a Markovian theory. The corresponding Langevin equation is then also linear in the sense that the dissipative parts (i.e., friction), remain linear in the particle’s velocity. To go beyond these model systems, a variety of nonlinear properties of complex fluids have been studied~\cite{Larson1999-yj}. An important example concerns shearing of complex fluids~\cite{gutsche2008colloids, Wilson2011-ip, Leitmann2013-wo, Harrer2012-rs, Winter2012-ja, Gomez-Solano2014-jh, Gazuz2009-er, Squires2005-mc, Benichou2013-yn}, which, experimentally, involves macroscopic rheometers~\cite{Isa2007-jo, Besseling2007-id, Weiss1999-fp, Smith2007-ac}. In such investigations, many nonlinear features have been observed, e.g., shear-thinning or thickening, the phenomenon of yielding, or the multifaceted nonlinear response to oscillatory shear. Following these macroscopic studies, the development of microrheology, where a single microscopic probe particle is driven through a (nonlinear) medium, allowed to investigate the fluids’ behavior at much smaller length scales~\cite{gutsche2008colloids, Wilson2011-ip, Leitmann2013-wo, Harrer2012-rs, Winter2012-ja, Gomez-Solano2014-jh, Gazuz2009-er, Squires2005-mc, Benichou2013-yn}. In  related experimental and theoretical studies, several  consequences of the nonlinearity of the bath have been reported. For example, it has been observed that the driven probe experiences an effective temperature which differs from the true bath temperature~\cite{Wilson2011-ip, Demery2019driven}, that it shows superdiffusive behavior~\cite{Benichou2013-yn, Winter2012-ja}, or shear thinning~\cite{Wilson2011-ip, Squires2005-mc}. More recently, experiments reported the occurrence of oscillatory modes~\cite{Berner2018-kk}, which are seen in a regime of linear, e.g., Newtonian-like behavior. The oscillatory dynamics in Ref.~\cite{Berner2018-kk} was reproduced using a generalized Langevin equation with negative memory at long times, which can induce persistent motion \cite{Zausch08, mitterwallner2020negative} and stress overshoots \cite{Fuchs03}. Furthermore, it was noted that the non-linear properties of a bath can already be detected in equilibrium~\cite{Muller2020-mo}. For example, in a nonlinear bath, the effective friction memory kernel may depend on the external potential~\cite{Daldrop2017external, Zaccone2018generalized, Kowalik2019memory, Lisy2019generalized, Muller2020-mo, Muller2019thesis, Tothova2021brownian}. These findings illustrate the difficulty to match micro and macro-rheology measurement, and make the study of non-Markovian baths all the more important. 

Here we study the motion of a particle trapped in a viscoelastic fluid, for different driving velocities, where, compared to previous work~\cite{Berner2018-kk}, we extend the experimentally accessible regime towards smaller velocities. We observe the previously reported linear regime, where the flow curve shows a plateau. For even smaller velocities, however, the viscosity is seen to increase in a pronounced manner, so that two distinct plateaus and two shear thinning processes are found. Theoretically, we reproduce this behavior by use of a previously introduced stochastic Prandtl Tomlinson (SPT) model \cite{Muller2019thesis, Muller2020-mo}. This model also reproduces the previously observed  oscillations~\cite{Muller2019thesis, Berner2018-kk} for the given shear rates. The model implies that each shear thinning process corresponds to a de-equilibration of an important set of bath degrees of freedom, and it allows to estimate the important length scales involved in these degrees.


\section{Experimental Details }
Our experiments were performed in an equimolar solution of the surfactant cetylpyridinium chloride monohydrate (CPyCl) and sodium salicylate (NaSal) in deionized water at a concentration of \SI{7}{\milli\Molar} and at constant temperatures of \SI{25}{\celsius} and \SI{30}{\celsius}, respectively. Under such conditions, the mixture forms an entangled viscoelastic network of worm-like micelles~\cite{Cates1990-lk}, with a structural relaxation time on the order $\tau_s=\SI{2.5}{\s}$ as determined by microscopic recoil experiments~\cite{Gomez-Solano2015-qu}. Typical length scales of worm-like micelles are between 100 and \SI{1000}{\nano\m}~\cite{Walker2001rheology}, and their typical mesh size is on the order of \SI{25}{\nano\m}~\cite{Buchanan2005-dd}. A small amount of silica particles with $2R= \SI{2.73}{\micro\m}$ diameter has been added to the fluid and a single particle has been optically trapped by a focused laser beam of wavelength of \SI{1064}{\nano\m}. This creates a harmonic potential with trap stiffness $\kappa$. The latter is determined from the equilibrium probability distribution of the particle in equilibrium, see e.g. Ref.~\cite{Berner2018-kk}. To probe the micro-rheological properties, the sample is translated with constant velocity $-v_0$ relative to the static optical trap by a piezo-driven stage with $v_0$ between 4 and \SI{400}{\nano\m\per\s}. Note that this driving is equivalent to a trap which is moving at velocity $v_0$ with respect to a fluid at rest. Overall it leads to a local shear rate of $v_0/2R$ close to the particle, and a Weissenberg number of $Wi=v_{0} \tau_{s}/2R$. The so-obtained Weissenberg numbers range between a few permille to almost unity in our experiments. We thereby extend the measurements by one decade towards smaller Weissenberg numbers compared to Ref.~\cite{Berner2018-kk}.
\begin{figure}[ht!]
		\includegraphics[scale=0.3]{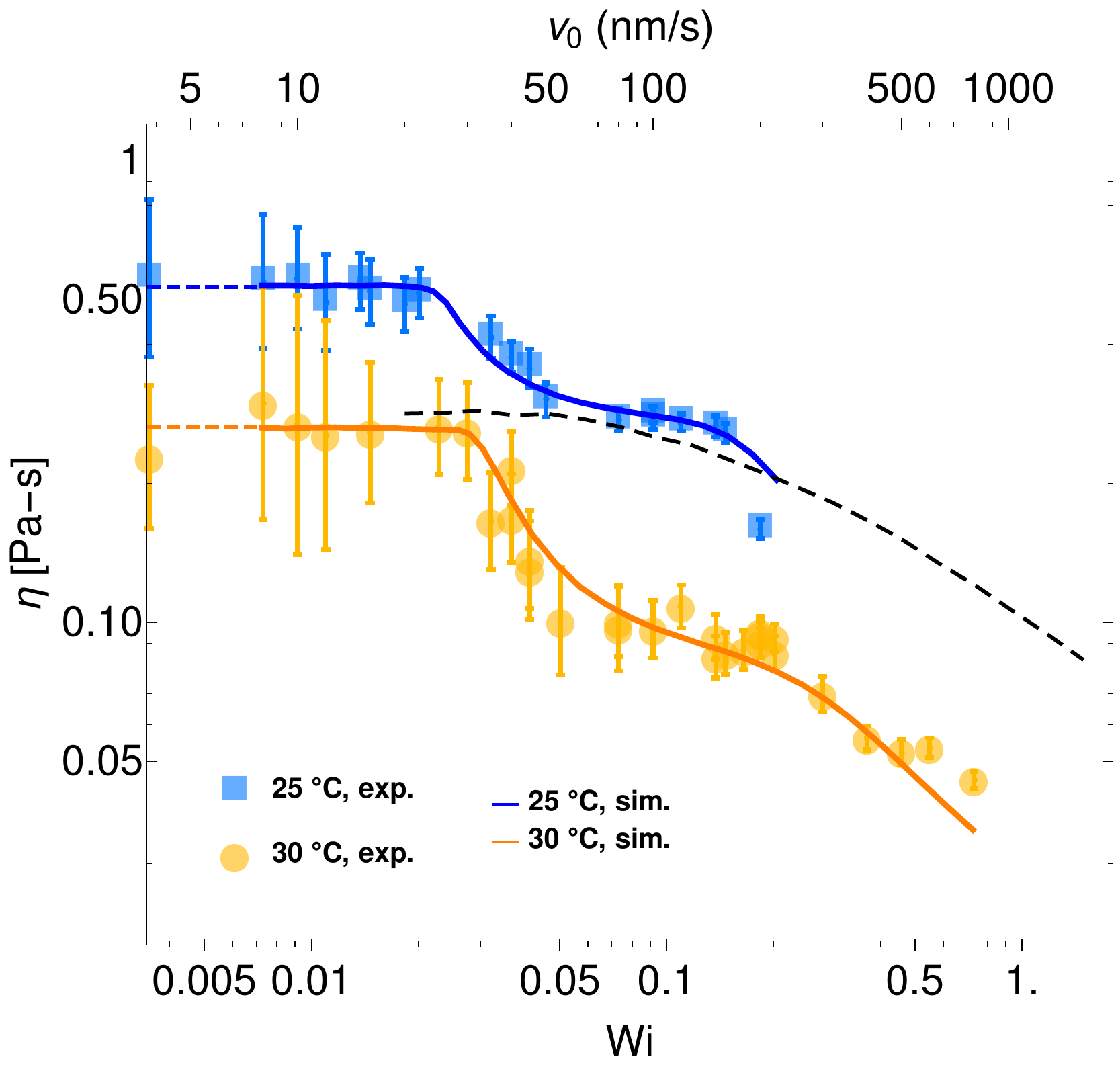}
	\caption{Microviscosity as a function of the driving speed (flowcurves). Symbols correspond to experimental data obtained using Eq.~\eqref{eqn:Flowcurve} for \SI{25}{\celsius} (blue squares) and \SI{30}{\celsius} (orange circles). Solid colored lines correspond to simulations using the stochastic PT model. Both temperatures exhibit the same two plateau shapes. For $v_0>\SI{200}{\nano\m\per\s}$, we recover the shear thinning behavior which is expected from macro-rheological measurements at \SI{25}{\celsius}, shown as a black dashed line. However, for $v_0<\SI{200}{\nano\m\per\s}$, the micro-viscosity further increases towards another plateau reaching values more than twice as high as the intermediate plateau. Dashed lines and data points on the $y$-axis give linear response values from simulation and experiment, respectively, using Eq.~\eqref{LinearResponseResult}. }
	\label{fig:flowcurve}
\end{figure}

When driving the particle with constant velocity through the fluid, it experiences a drag force which leads to a displacement relative to the trap center. Using Stokes law and the trap stiffness, we can measure the velocity-dependent microviscosity, given by
\begin{equation}
	\eta = \frac{\kappa}{6\pi R v_{0}}\, \left\langle x(t) - v_{0}t\right\rangle . \label{eqn:Flowcurve}
\end{equation}
Here, $\left\langle x(t) - v_{0}t\right\rangle$ directly corresponds to the average position of the particle, relative to the center of the trap. Note that no mean displacement in the direction orthogonal to the driving is observed.

The resulting flow curves are shown in Fig.~\ref{fig:flowcurve}, for two temperatures \SI{25}{\celsius} (blue squares), and \SI{30}{\celsius} (orange circles). Each data point corresponds to a single experiment of $\sim \SI{1000}{s}$, which is long enough for the particle to fully explore the trap. Errorbars are directly obtained from the standard deviation of the particle's position. Both curves exhibit the same trend. For driving speeds approaching Weissenberg numbers of order unity, the viscosity shows shear thinning, as expected. This is also confirmed by the macro-viscosity which is also shown in Fig.~\ref{fig:flowcurve} for the temperature of \SI{25}{\celsius} for comparison (black dashed line, obtained by a plane-plane rheometer). Indeed, for that temperature, the data corresponding to micro and macro viscosities agree reasonably well for $v_0>\SI{50}{\nano\m\per\s}$. We note that this data range was available in Ref.~\cite{Berner2018-kk}. From this close resemblance, together with the fact that the Weissenberg number is of the order of only $5$\% for $v_0=\SI{50}{\nano\m\per\s}$, at first glance one might conclude that these velocities are in the regime of linear response.

In contrast to this interpretation, however, the observation of oscillations in the so called mean conditional displacements (see details below), suggests that the system is far from equilibrium even at the small driving speed~\cite{Berner2018-kk} of $v_0=\SI{50}{\nano\m\per\s}$. When extending the range of velocities to smaller values, an astonishing observation is made: The viscosity shows a pronounced second shear thinning transition, connected to another plateau in the flow curve, of more than twice the viscosity value compared to the second plateau. The corresponding shear thinning process sets in at a Weissenberg number of roughly $2$\%, implying that this dimensionless quantity is not useful to understand that process. This conjecture is underpinned by the observation that the macro-viscosity seems to not display the shear thinning at the mentioned small velocities.
\begin{figure*}[ht!]
	\centering
	\includegraphics[width=.6\textwidth]{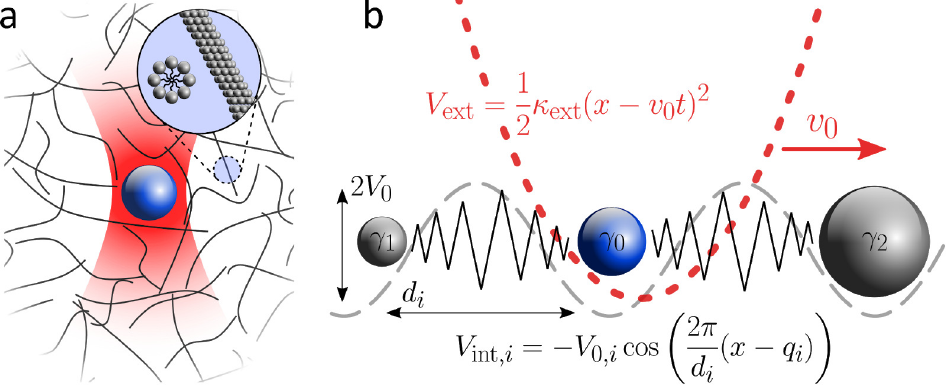}
	\caption{\small{\textbf{a.} Sketch of the experimental system, highlighting the probe suspended in a worm-like micellar fluid. The particle is trapped with an optical tweezer, using a highly focused laser beam to form a harmonic potential for the particle. \textbf{b.} Sketch of the Stochastic Prandtl-Tomlinson model. The tracer particle (bare friction coefficient $\gamma_0$) is subject to a harmonic potential $V_\textrm{ext}$ with stiffness $\kappa$ , moving at constant speed $v_0$. The tracer is also linked to several (here two) bath particles with friction coefficients $\gamma_1$ and $\gamma_2$, respectively, through periodic interaction potentials $V_{\textrm{int},i}$.}}
	\label{fig:ModelSketch}
\end{figure*}

\section{Theory and model}
To rationalize these but also previous experimental observations~\cite{Muller2020-mo}, we consider a simple model where the tracer particle is coupled to a small number of bath particles which mimic the fluid environment (see sketch of Fig.~\ref{fig:ModelSketch}). As has been realized before \cite{Zwanzig2001-bd, caldeira1981influence},  using models with bath particles yields visco-elastic bath behavior, where linear bath-tracer interactions result in exactly solvable models \cite{Zwanzig2001-bd, caldeira1981influence,Muller2020-mo}. As analyzed in Ref.~\cite{Muller2020-mo}, many features of the considered micellar suspension are due to the nonlinearity of the fluid, for which finding Langevin equations poses a general theoretical challenge \cite{zwanzig1973nonlinear,Kruger_2016,Meyer17}. In particular, an anharmonic interaction potential $V_\textrm{int}$ (specified below) is required. Moreover, as argued in Ref.~\cite{Muller2020-mo}, the potential $V_\textrm{int}$ needs to be unbounded in the sense that it is finite for any particle separation. This allows the bath and tracer particles to be arbitrarily far away from each other. Within this model, e.g. the effect of shear thinning is reproduced, because tracer and bath particles are able to move at different (average) speeds. The simplest potential $V_\textrm{int}$ fulfilling these requirements  is a sinusoidal function. In Ref.~\cite{Muller2020-mo}, using this one dimensional model with just a single bath particle was found to indeed well capture important features of the considered three dimensional system {\it in equilibrium}, such as a dependence of the resulting memory kernel on the stiffness of the external potential \cite{Daldrop2017external}. Here, in the driven case, we saw the need to extend this model to include more than one bath particle (detailed below), so that, for bath particle $i$,
\begin{equation}
	V_{\textrm{int},i} = V_{0,i}\, \cos\left(\frac{2\pi}{d_{i}}(x-q_{i})\right) \label{InteractionPotential}
\end{equation}
where $V_{0,i}$ is the amplitude and $d_i$ is a length scale corresponding to bath particle $i$. $x$ and $q_i$ are the position coordinates of tracer and $i$-th bath particle, respectively. The parameter $d_i$ in Eq.~(\ref{InteractionPotential}) mimics an important length scale in the micellar bath, see above for typical numbers. In addition, the tracer particle is subjected to a harmonic potential of stiffness which results from the laser tweezer. As the tweezer is moving with velocity $v_0$ relative to the fluid, the trap potential $V_\textrm{ext}$ reads,
\begin{equation}
	V_\textrm{ext} = \frac{1}{2}\kappa \left(x - v_{0}t\right)^{2}  .\label{ExternalPotential}
\end{equation}
Accordingly, the equations of motion (with $N$ bath particles) are given by
\begin{eqnarray}
	\gamma_{0}\dot{x}(t) &=& -\kappa \left(x - v_{0}t\right) - \sum_{i=1}^{N}\frac{2\pi}{d_{i}}\, V_{0,i}\, \sin\left(\frac{2\pi}{d_{i}}(x-q_{i})\right) \nonumber \\
	&&\,\, + \,\, \xi_{0}(t),  \label{LangevinEquationTracer} \\
	\gamma_{i}\dot{q}_{i}(t) &=& \frac{2\pi}{d_{i}}\, V_{0,i}\, \sin\left(\frac{2\pi}{d_{i}}(x-q_{i})\right) + \xi_{i}(t) . \label{LangevinEquationBath}
\end{eqnarray}
Here, $\gamma_0$ and $\gamma_i$ are the bare friction coefficients of tracer and bath particle $i$, respectively. These coefficients are linked to the corresponding random forces $\xi_0$ and $\xi_i$ via the following standard properties,
\begin{equation}
	\left\langle \xi_{i}(t) \right\rangle = 0 \quad\text{and}\quad \left\langle \xi_{i}(t)\xi_{j}(t') \right\rangle = 2k_{B}T\gamma_{i}\delta_{ij}\delta(t-t')  .  \label{NoiseCorrelation}
\end{equation}
We assume in the following that the driving started at an infinite time in the past, so that the system is in a steady state for $t>0$. Practically, this means that from experimental as well as simulated trajectories, the initial parts, corresponding to equilibration, are removed.

The interaction potential of Eq.~(\ref{InteractionPotential}) makes this model reminiscent of the so-called Prandtl-Tomlinson (PT) model, which is used to study dry friction~\cite{Prandtl1928-yd, Tomlinson1929-pj}. Eqs.~(\ref{LangevinEquationTracer}) and (\ref{LangevinEquationBath}) extend this model to allow the background (our bath particles) to be stochastic and dynamic, and to contain several bath particles. This Stochastic Prandtl Tomlinson (SPT) model~\cite{Muller2020-mo, Muller2019thesis} thus reduces to the original PT model when setting $N=1$ and letting $\gamma_1$ approach infinity, so that the bath particle becomes a stationary background potential. This difference is quite intuitive, as our micellar background is dynamic, while the potentials considered in dry friction are rather static. Note that the notion 'Stochastic Prandtl-Tomlinson model' has been used for other models which are different from the one used here~\cite{Jagla2018-ti, Van_Spengen2010-yu}.

The solid curves shown in Fig.~\ref{fig:flowcurve} give the outcomes of the model of Eqs.~(\ref{LangevinEquationTracer}) and (\ref{LangevinEquationBath}) (see Table \ref{table:parameters2baths} for parameters used), using Eq.~(\ref{eqn:Flowcurve}). For small velocities $v_0$, we observe a linear response regime, where the micro-viscosity is independent of $v_0$. In this regime, the viscosity can be obtained also via linear response (see also Ref.~\cite{Muller2020-mo})
\begin{equation}
   \lim\limits_{v_{0}\rightarrow0} \gamma(v_{0}) = \frac{\kappa^{2}}{k_{B}T}\,\int_{0}^{\infty} dt\, \left\langle x(t)x(0)\right\rangle_{eq},  \label{LinearResponseResult}
\end{equation}
shown as the horizontal dashed lines in the graph. It is insightful to discuss the case of high potential barriers $V_{0,i}/k_{B}T$, which appear appropriate to fit the experimental data (see Table \ref{table:parameters2baths}). Then the regime of small driving speeds corresponds to the case where all particles move (approximately) with the same average velocity $v_0$. Because of this, the microviscosity in Eq.~(\ref{eqn:Flowcurve}) can be found to a good approximation from $\eta \approx\frac{1}{6\pi R} \,\left(\gamma_{0}+\sum_{i}\gamma_{i}\right)$. In the graph, the resulting value is not shown, as it is indistinguishable from the dashed lines. Experimentally, this corresponds to the case where the slow colloidal particle drags its surrounding with it, and thus feels a large friction.

The interaction potential $V_{\textrm{int},i}$ in Eq.~(\ref{InteractionPotential}) supports a maximal force of $\frac{2\pi}{d_{i}}\,V_{0,i}$. If the force between the tracer and the bath particle exceeds that value, the structure breaks, and the velocity of the bath particle is (on average) smaller than $v_0$: As a result, the system shows a shear thinning behavior. The critical velocity where this happens can be estimated by balancing the mentioned maximal force with the drag force $\gamma_{i}v_{0}$ of particle $i$, yielding (in absence of noise) $v_{i,cr}\approx \frac{2\pi}{\gamma_{i}d_{i}}\,V_{0,i}$ . For the curves in Fig.~\ref{fig:flowcurve}, we use two bath particles, with distinct critical velocities, resulting in the two-plateau structure seen in the graph. As mentioned, for very small velocities, both bath particles move (approximately) with the same velocity as the tracer. The smaller critical velocity corresponds to the “larger” bath particle, i.e., the one with a larger value of $d_i$. Using the values of Table \ref{table:parameters2baths}, we estimate this velocity to be $27$ and $\SI{36}{\nano\m\per\s}$ for the two temperatures, respectively, which fits well to the numerical curves. Beyond this velocity, the viscosity decreases towards the second plateau. On the regime of the second plateau, the larger particle is thus far from equilibrium, while the second (“smaller”) bath particle is still close to equilibrium. Once the second critical velocity (estimated to $321$ and $\SI{655}{\nano\m\per\s}$, respectively) is reached, also the second particle starts shear thinning. The model thus implies the interpretation that two distinct sets of degrees of freedom of the bath display very different critical velocities, yielding an intermediate state with half of them out of equilibrium. Notably, the often employed notion of fast and slow degrees of freedom~\cite{Zwanzig2001-bd} is demonstrated here explicitly, at least in our model.

\begin{table}
	\centering 
	\begin{tabular}{|c|c|c|c|}
		\hline
		\hline
		\,$T$ \, &  \,$(\gamma_{0},\gamma_{1},\gamma_{2})$ \, &  \,$(V_{0,1}, V_{0,2})$ \, &  \,$(d_{1}, d_{2})$\, \\ 
		\, [\SI{}{\celsius}]\, &   \, [\SI{}{\micro\N\s\per\m}]\, &  \, $[k_{B}T]$\, &  \, [\SI{}{\nano\m}]\, \\ 
		\hline
		$25$  &   $(0.06, 1.5, 1.5)$  &  $(7.6, 24.0)$  &  $(400, 15000)$ \\
		\hline
		$30$  &    $(0.04, 0.5, 1.0)$  &  $(4.0, 20.6)$  &  $(320, 15000)$ \\
		\hline
		\hline
	\end{tabular}
\caption{Parameters of the SPT model used to simulate the flow-curves  in Fig.~\ref{fig:flowcurve} and the mean conditional displacements in Fig.~\ref{fig:MCDPlots}, for the two temperatures. In both experiment and simulation, the values of trap stiffness for the flow curve and MCDs are $\kappa=\SI{1.5}{\micro\N\per\m}$ and $\kappa=\SI{0.3}{\micro\N\per\m}$, respectively.}
\label{table:parameters2baths}
\end{table}

\begin{figure*}
	\begin{minipage}{0.45\textwidth} \textbf{\large{(a)}} 
    \end{minipage}
    	\hfill
    \begin{minipage}{0.45\textwidth} \textbf{\large{(b)}} 
    \end{minipage}  \\
    
    \includegraphics[width=.5\textwidth]{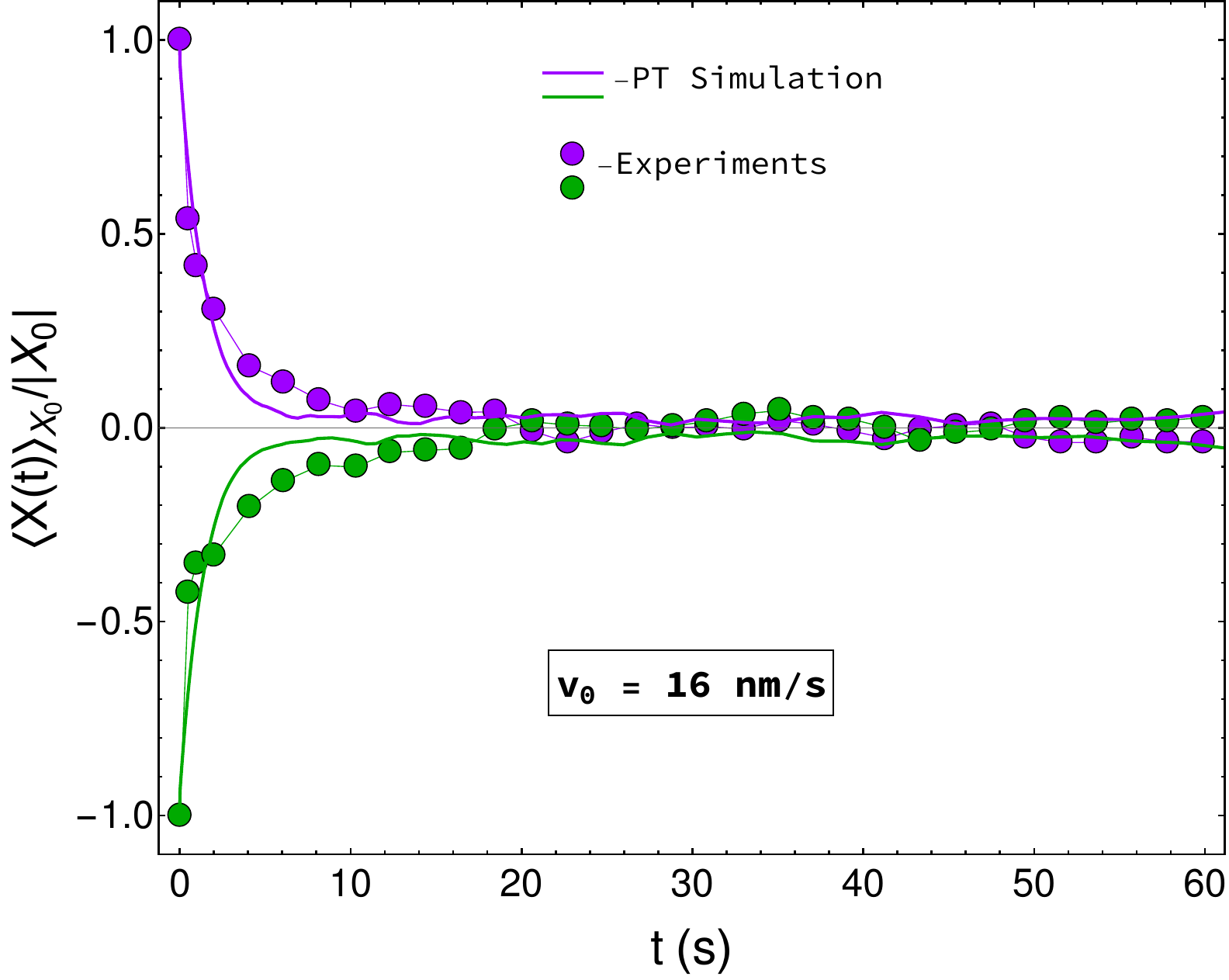}
        \hfill
    \includegraphics[width=.47\textwidth]{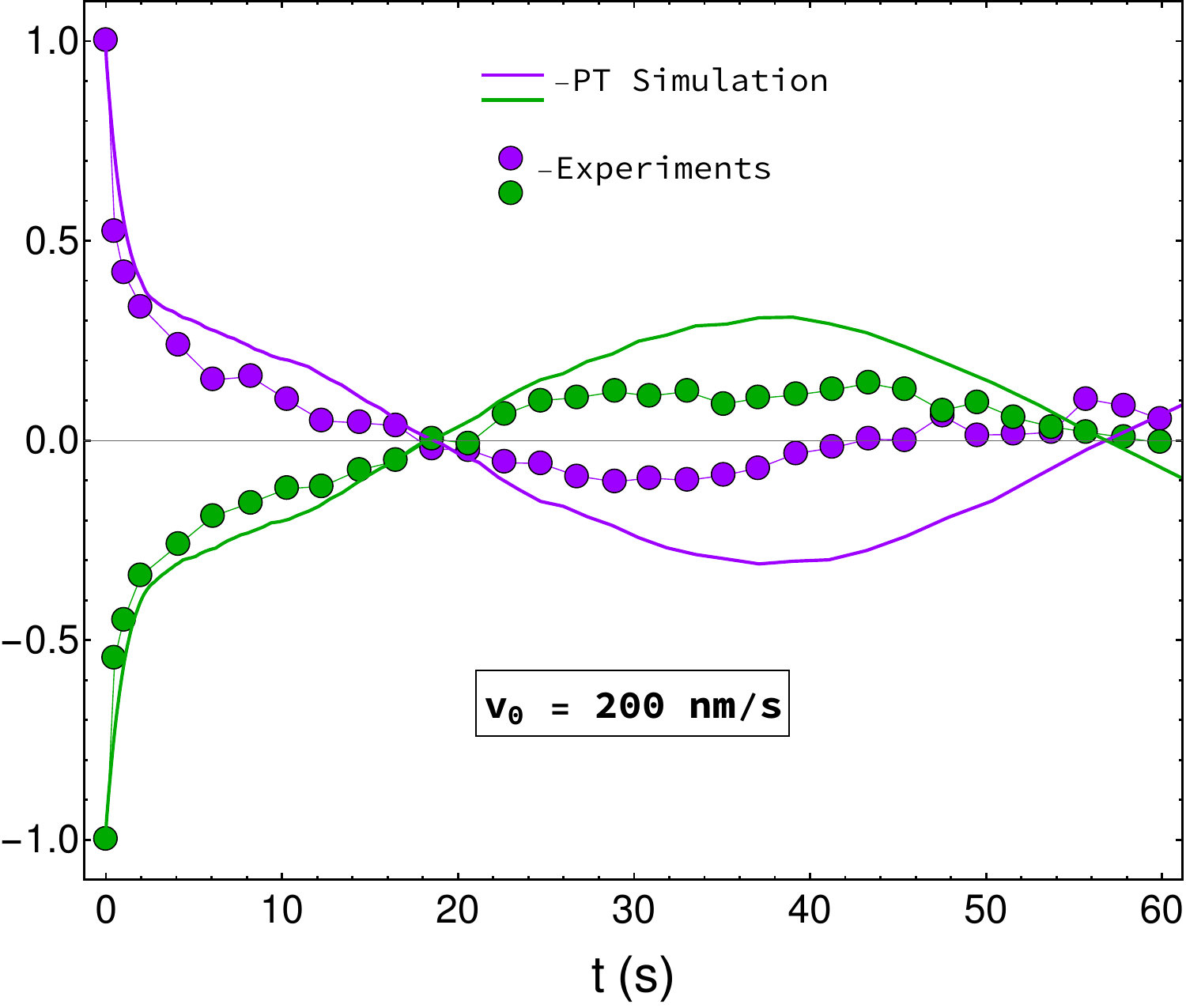}
    \caption{Mean conditional displacements (a) in the linear response regime, $\SI{16}{\nano\m\per\s}$ and (b) in the regime of intermediate plateau, $\SI{200}{\nano\m\per\s}$. Circles correspond to experiments (at \SI{30}{\celsius}), and lines correspond to the Prandtl-Tomlinson simulations. While we recover previously observed oscillations~\cite{Berner2018-kk} for $\SI{200}{\nano\m\per\s}$, the linear regime  (left, $\SI{16}{\nano\m\per\s}$) exhibits none. As shown in Appendix \ref{sec:3B}, adding more bath particles in the PT model leads to decoherence,  decreases the amplitude of the oscillations, and improves agreement with experiments.}
\label{fig:MCDPlots}
\end{figure*}

The flow curve concerns the average position $\left\langle x - v_0 t\right\rangle $ of the particle, as seen from Eq.~(\ref{eqn:Flowcurve}), and we now aim to address the particle’s fluctuations. Subtracting the particle’s mean position, i.e. using $X(t)=x(t)-\left\langle x - v_0 t\right\rangle $, yields a stochastic variable $X$ with zero mean. Its fluctuations can be quantified using the so-called Mean Conditional Displacement (MCD)~\cite{Berner2018-kk}. It is defined by $\left\langle X\right\rangle_{X_0} = \int_{-\infty}^{\infty}dX\, X\, P(X,t|X_0,0)$, with $P(X,t|X_0,0)$ the  probability for the particle position $X$ under the condition that the initial position at time $t=0$ is $X_0$. In previous work, we observed that MCDs show oscillations in this system~\cite{Berner2018-kk}, which were observed at a temperature of \SI{25}{\celsius}, for Weissenberg numbers in the range of $0.04$ to $0.34$. Can the SPT model also explain the occurrence of oscillations? To address this question, we show in Fig.~\ref{fig:MCDPlots} experimental and simulated MCD curves, this time for a temperature of \SI{30}{\celsius}, to also emphasize the same phenomenology at the two temperatures. The MCD curves generally are found to a good approximation linear in $X_0$ (both in experimental as well as in simulated data, see also Fig.~7 of Ref.~\cite{Berner2018-kk}), so that division by $X_0$ yields the shown $X_0$-independent curves. The figure distinguishes between positive and negative values of $X_0$, yielding the positive and negative flanks shown. The observed mirror symmetry of the data with positive and negative $X_0$ further emphasizes the linearity in $X_0$. Fig.~\ref{fig:MCDPlots}a) shows the MCD curve for a small velocity of $v_0=\SI{16}{\nano\m\per\s}$, which corresponds to the regime of linear response in Fig.~\ref{fig:flowcurve}. For this velocity, the MCD curve decays monotonically to zero, as expected near equilibrium~\cite{Berner2018-kk}, and also in agreement with the SPT model. Fig.~\ref{fig:MCDPlots}b) shows the MCD curve for a larger velocity of $v_0=\SI{200}{\nano\m\per\s}$, a velocity placed on the intermediate plateau in the flow curve in Fig.~\ref{fig:flowcurve}, thus in the regime corresponding to the curves shown in Ref.~\cite{Berner2018-kk, Muller2019thesis}. This curve shows pronounced oscillations, which are reproduced by the SPT model. How can these be understood? As described above, the intermediate plateau is beyond the critical velocity of the larger bath particle, which is thus far from equilibrium. It thus moves with an average speed much smaller than $v_0$. It can for the sake of argument be assumed to stand still, so that the tracer is moving in a stationary periodic potential. It is thus subject to a periodic force, which results in the seen oscillations. The period of oscillations is in this approximation given by $d/v_0$, which matches well the one observed in Fig.~\ref{fig:MCDPlots}b). In Ref.~\cite{Berner2018-kk}, the frequency of oscillations was indeed found to scale linearly with $v_0$, an observation which can now be understood. This analysis thus identifies an important length scale in the system, of the order of \SI{15}{\micro\m}. Passing over spatial variations on that scale (which are almost stationary as seen from the colloidal particle) seems to cause the observed oscillations.

The amplitude of oscillations is larger in the SPT model as compared to experiments, which we attribute to a number of idealizations of the model. For example, the background potential of the SPT model is perfectly periodic with a sharp length scale $d_i$. A real micellar solution, however, will exhibit a range of length scales, which naturally leads to decoherence. Indeed adding more bath particles, with slightly different parameters (see Appendix \ref{sec:3B}) leads to a loss of coherence, and the amplitude of the resulting oscillations is reduced. Notably, adding more bath particles with appropriate parameters keeps the flow curve unaltered, but changes the MCD. This confirms the expectation that flow curve and MCD are not one to one related. The flow curve concerns the mean motion, and the MCD quantifies fluctuations.

In Ref.~\cite{Muller2020-mo}, a single bath particle was found to be sufficient to capture experimental observations in the SPT´ model. It is natural that a system close to equilibrium (as in Ref.~\cite{Muller2020-mo}) is easier to model compared to a system far from equilibrium, as addressed here. Indeed, an open question is whether the larger length scale  of $\sim\SI{15}{\micro\m}$ can be detected in equilibrium.

Finally, we note that MCD curves and the flow curve can  be modeled quantitatively using the same parameters (see Table \ref{table:parameters2baths}), with one exception: We did not succeed in obtaining the correct amplitude of the flow curve in this procedure. We therefore allowed the amplitude of flowcurve to be multiplied by a free parameter in our simulations, which turned out to be $\SI{4.5}{}$. We attribute this difficulty to the different values of trap stiffness used for flow curve and MCDs of $\SI{1.5}{\micro\N\per\m}$ and $\SI{0.3}{\micro\N\per\m}$, respectively, which were used for experimental reasons. This makes fitting both curves simultaneously even more challenging.

\section{Conclusion}
In this work we describe micro-rheological experiments where a single colloid is driven by an optical tweezer within a viscoelastic fluid. The presented fluid, a micellar solution, shows a flow curve with two distinct shear thinning regimes, with an intermediate plateau in between. Theoretical modeling via a stochastic Prandtl Tomlinson model captures the observed behavior, and implies that, on the intermediate plateau, one set of bath degrees of freedom is far from equilibrium, while another set is still in equilibrium. The intermediate plateau corresponds to the linear response regime of macro-rheology, so that the shear thinning process at even smaller driving velocities is a purely microscopic effect, and can thus easily be overlooked. The mean conditional displacements show oscillations on the intermediate plateau, which are also reproduced in the theoretical model. The oscillations allow extraction of a length scale, which is as large as \SI{15}{\micro\m}, and whose nature and origin have to be investigated in future work. Theoretically, this could be based on approaches using density functional theory \cite{Penna04, Rauscher07}, where changes in fluid density due to a driven tracer, and their length scales, can be investigated.

\section{Acknowledgments}
The authors thank Boris M\"{u}ller for discussions at the early stages of this work. The theoretical parts of this work build on the initial findings provided in his thesis~\cite{Muller2019thesis}. RJ and MK are also thankful to Marcus M\"{u}ller for insightful discussions and suggesting to look at Ref.~\cite{Rauscher07}. FG and JB thank Jakob Steindl for the micellar solution and the macro-rehology measurements.\\
\textbf{Funding}: FG acknowledges the support by the Alexander von Humboldt foundation. This project is funded  by the Deutsche Forschungsgemeinschaft (DFG, German Research Foundation), Grant No. SFB 1432 - Project ID 425217212.  RJ acknowledges the support by the G\"{o}ttingen Campus QPlus program.\\ 
\textbf{Competing interest}: The authors declare no competing interest.

\section{Data availability}
The data that support the findings of this study are available from the corresponding author upon reasonable request.

\appendix
\section{Including more bath particles}\label{sec:3B} 
\begin{figure}[ht!]
	\centering
	\includegraphics[scale=0.3]{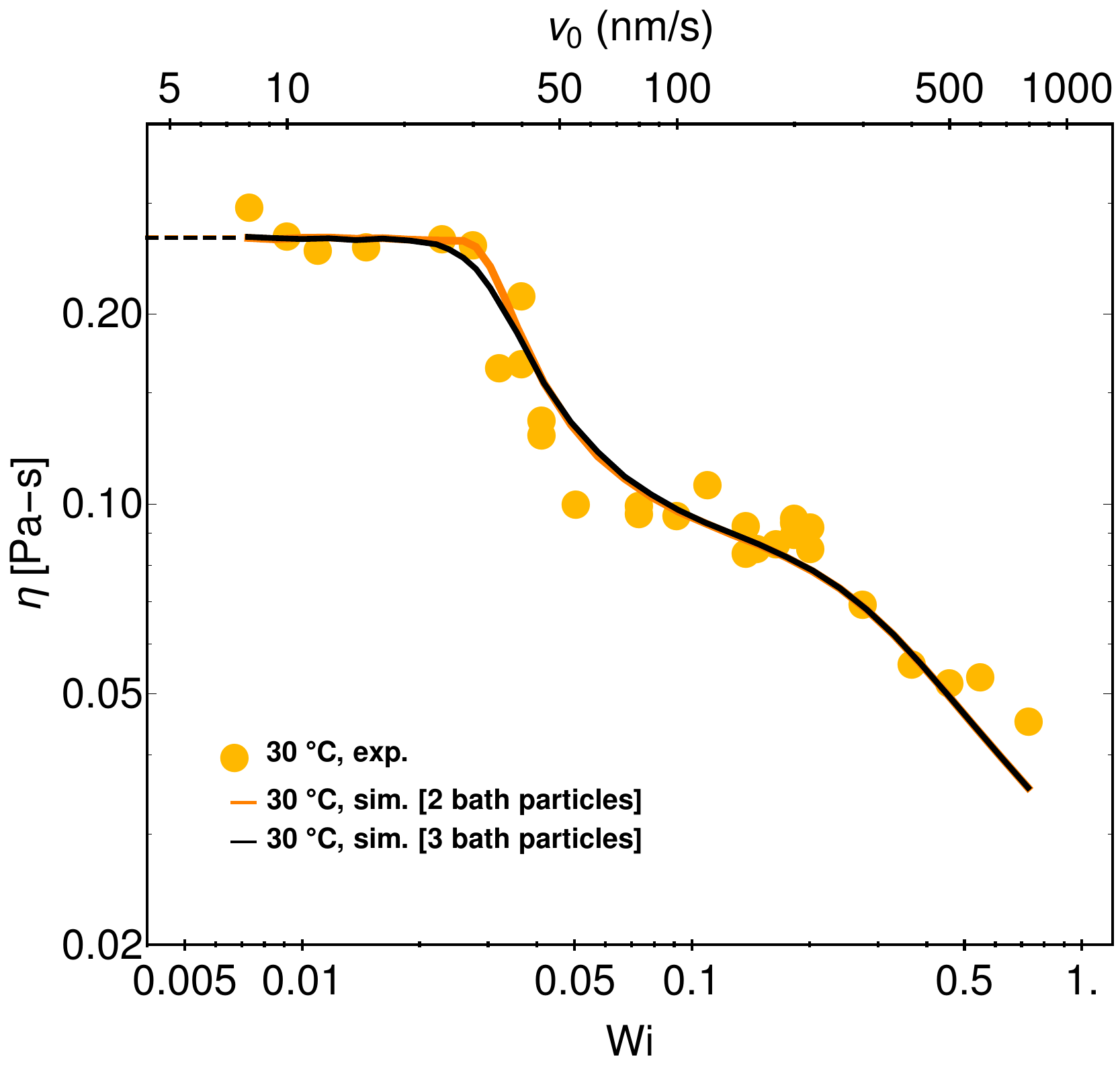}
	\caption{\small{Comparison of flowcurves measured at \SI{30}{\celsius} to simulations with two and three bath particles for a trap-stiffness, $\kappa = \SI{1.5}{\micro\N\per\m}$. Circles correspond to experiments and lines correspond to the Prandtl-Tomlinson simulations (orange and black lines for $2$ and $3$ bath particles, respectively).}}
	\label{fig:FlowcurveEffectOfParticleNumber}
\end{figure}

\begin{figure}[ht!]
    \centering
         \includegraphics[scale=0.3]{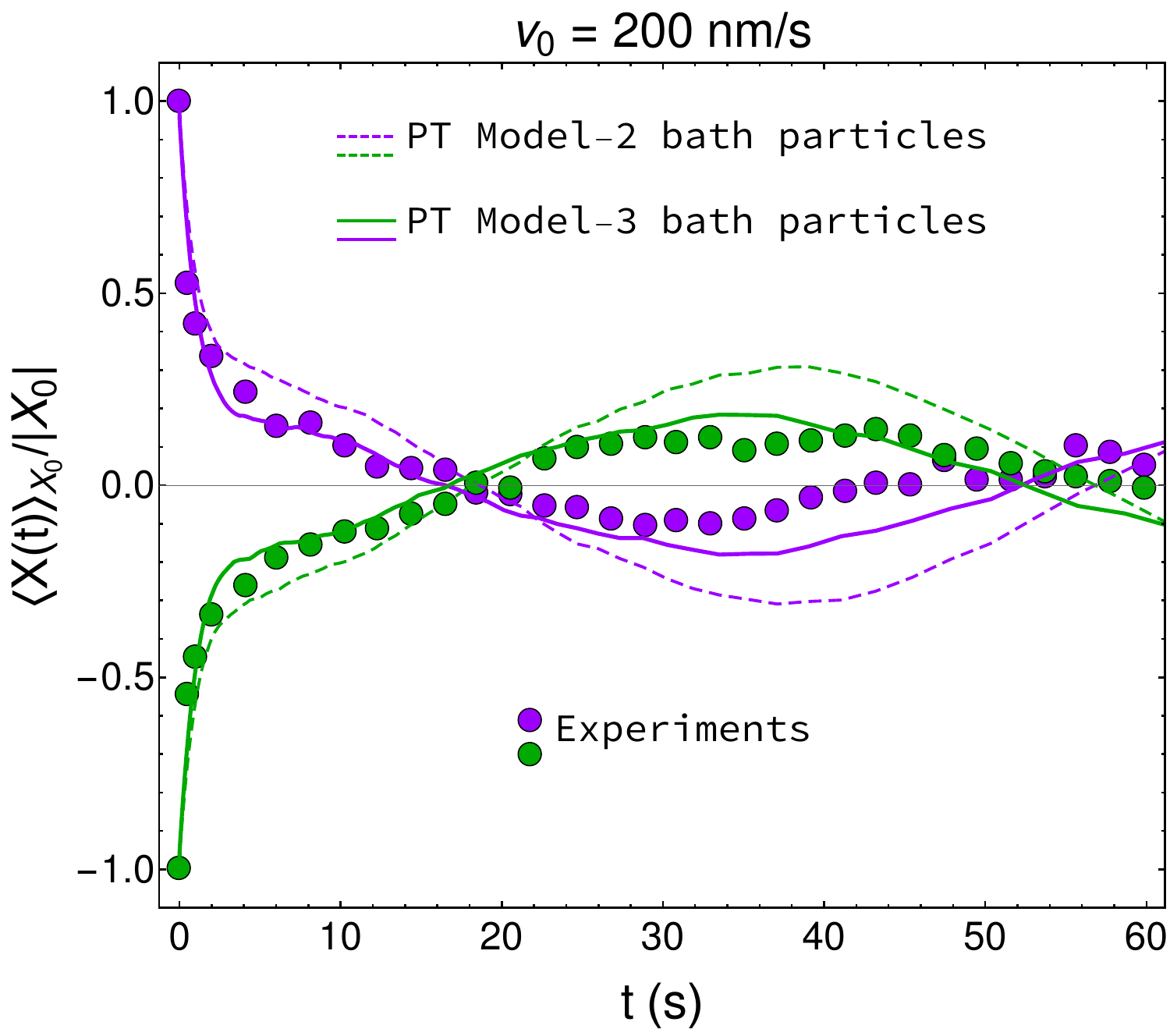}
    \caption{Comparison of MCD curves measured at \SI{30}{\celsius} with simulations with two and three bath particles, for a driving velocity, $v_{0}=\SI{200}{\nano\m\per\s}$ and trap-stiffness, $\kappa = \SI{0.3}{\micro\N\per\m}$. Circles correspond to experiments and lines correspond to the Prandtl-Tomlinson simulations (dashed and solid lines for $2$ and $3$ bath particles, respectively). Adding more bath particles leads to de-coherence, resulting in the reduction of amplitude of the oscillations.}
\label{fig:MCDPlots_comparing-2_vs_3_baths}
\end{figure}

As discussed in the main text, including more bath particles into the model leads to de-coherence, resulting into the reduction in amplitude of oscillations in the MCD. We choose the parameters for the third bath particle (see Table \ref{table:parameters3baths30T}) in such a way that the flow curves remain unaffected.
\begin{table}[ht!]
	\centering 
	\begin{tabular}{|c|c|c|}
		\hline
		\hline
		 \,$(\gamma_{0},\gamma_{1},\gamma_{2},\gamma_{3})$ \, &  \,$(V_{0,1}, V_{0,2}, V_{0,3})$ \, &  \,$(d_{1}, d_{2}, d_{3})$\, \\ 
		 \, [\SI{}{\micro\N\s\per\m}]\, &  \, $[k_{B}T]$\, &  \, [\SI{}{\nano\m}]\, \\ 
		\hline
		 $(0.04, 0.5, 0.5, 0.5)$  & $(4.0, 8.2, 10.8)$  &  $(320, 12000, 15000)$ \\
		\hline
		\hline
	\end{tabular}
	\caption{\small{Parameters used to simulate the MCD curves and flow curve with three bath particles at temperature of \SI{30}{\celsius}. In both experiment and simulation, the values of trap stiffness for the flow curve and MCDs are $\SI{1.5}{\micro\N\per\m}$ and $\SI{0.3}{\micro\N\per\m}$, respectively.}}
	\label{table:parameters3baths30T}
\end{table}
In Fig.~\ref{fig:FlowcurveEffectOfParticleNumber}, we show the flowcurves generated with the parameters of Table \ref{table:parameters2baths}, i.e with two bath particles and that of Table \ref{table:parameters3baths30T}, i.e. with three bath particles, and our experimental data. Likewise, in Fig.~\ref{fig:MCDPlots_comparing-2_vs_3_baths}, we show the MCD curve in the regime of intermediate plateau, i.e. $v_{0}=\SI{200}{\nano\m\per\s}$ computed with the parameters of Table \ref{table:parameters2baths} i.e. with two bath particles and those of Table \ref{table:parameters3baths30T} (i.e. with three bath particles) and the experimental data. We see from Fig.~\ref{fig:FlowcurveEffectOfParticleNumber} and Fig.~\ref{fig:MCDPlots_comparing-2_vs_3_baths} that while including additional bath particles in the SPT model may not change the flow curve, it results in reduction of amplitude of oscillations, leading to a better agreement with experiments.

\end{document}